# An Algorithm for Flow Control in Computer Networks Based in Discrete Control Theory

G. Millán, *Member, IEEE*

*Abstract*— Developing of an effective flow control algorithm to avoid congestion is a hot topic in computer network society. This document gives a mathematical model for general network at the beginning, and then discrete control theory is proposed as a key tool to design a new flow control algorithm to avoid congestion in the high-speed computer network, the proposed algorithm ensures stability of network system. The results of the simulation show that the proposed method can adjust the sending speed and the queue level in the buffer quickly and effectively. In addition, the method is easy to implement and apply to high-speed computer network.

## I. Introducción

LAS redes de computadoras de alta velocidad generalmente son redes troncales de almacenamiento y retransmisión que conmutan nodos y enlaces de comunicación sobre determinada topología, donde todos los nodos y enlaces se caracterizan por sus propias capacidades para almacenar y transmitir paquetes, respectivamente. Un nodo que alcanza su capacidad máxima de almacenamiento debido a la saturación de sus procesadores o a la de uno o más de sus enlaces de transmisión salientes se llama congestionado. Así, algunos de los paquetes que llegan a un nodo congestionado pueden no ser aceptados y deberán de ser retransmitidos en una instancia posterior, hecho que conduce al deterioro de los parámetros de rendimiento de las redes y, en el peor escenario, al colapso de la red. Por lo tanto, el control de congestión es un problema importante que surge de la gestión de las redes. Un esquema de control de flujo puede ajustar la tasa de envío de paquetes en el host de origen para evitar la congestión; por ello un esquema de control de flujo adecuado para una red es una forma directa de actuar sobre su rendimiento.

Muchos sistemas complejos, como es el caso de las redes de computadoras de alta velocidad, pueden ser analizados por la teoría clásica de control. Por ello, una cantidad creciente de investigaciones se dedica a fusionar la teoría de control con el control de flujo. La primera aplicación de la teoría de control al control de flujo apareció en las redes ATM. En [1]-[4] se propone el control de tráfico basado en la tasa disponible de bits (ABR). En estos esquemas, si la longitud de la cola en un conmutador es mayor que un umbral de aceptación, entonces se establece un dígito binario sobre la cabecera de la célula de control de admisión. Sin embargo, todos ellos sufren problemas de estabilidad al exhibir una dinámica oscilatoria y requerir una gran cantidad de buffer para evitar la pérdida de células. En consecuencia los algoritmos de tasa explícita son ampliamente considerados e investigados. En [5] se expone un compendio excelente al respecto. La mayoría de los esquemas de tasa explícita existentes no consideran dos partes fundamentales en el diseño del control retroalimentado: 1) El análisis de la dinámica de la red en bucle cerrado, 2) La interacción con el tráfico VBR. En [4] se propone un algoritmo de tasa explícita que calcula las tasas de entrada dividiendo el ancho de banda disponible por las conexiones activas. En [6] el problema de diseño del control se plantea como un problema estándar de rechazo de perturbaciones donde el ancho de banda disponible actúa como una perturbación para el sistema. En [7] el problema se formula en el dominio del control de tipo estocástico donde la perturbación se modela como un proceso autoregresivo; el nodo en cuestión tiene que, así, estimar este proceso utilizando mínimos cuadrados recursivos. En [8] y [9] se usa el principio de Smith para derivar un control en el caso de que un búfer FIFO se mantenga en los enlaces de salida.

En todos los trabajos anteriores el diseño del algoritmo de control encuentra su base en la teoría del sistema en tiempo continuo. Pero como se sabe el control en tiempo discreto es eficaz en el diseño asistido por computadora y fácil de poner en ejecución. Luego, en este trabajo se diseña un algoritmo de control discreto para satisfacer requisitos prácticos de control.

## II. Modelo Matemático Generalizado

En esta sección se desarrolla el modelo matemático de una red general conmutada con filosofía de servicio de paquetes de almacenamiento y reenvío; es decir, los paquetes entran en la red desde los nodos fuente ubicados en los bordes, y luego se almacenan y reenvían a lo largo de una secuencia compuesta de nodos intermedios y enlaces de comunicaciones, llegando finalmente a sus nodos de destino. La Fig. 1 representa una red de conmutación de paquetes de almacenamiento y reenvío. En la Fig. 1 $S_i$ ($i = 0, \cdots, n$) denota al nodo fuente; $D_i$ ($i = 0, \cdots, n$) denota al nodo destino; $B$ representa al nodo cuello de botella; $u_{0i}$ ($i = 0, \cdots, n$) representa la tasa de envío, y $q_i$ ($i = 0, \cdots, n$) denota el nivel de cola del buffer.

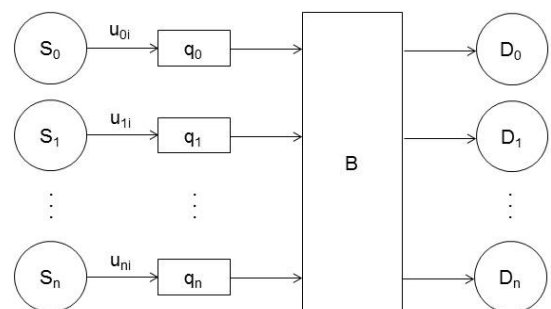

Fig.1. Red conmutada de paquetes con almacenamiento y reenvío.

G. Millán, Departamento de Ingeniería Eléctrica, Universidad de Santiago de Chile, ginno.millan@usach.cl.

Una aproximación al tamaño del buffer (o nivel de cola del buffer) es utilizar buffers de tamaño variable. La ventaja es un mejor uso de la memoria al costo de una gestión de buffer más complicada. Otra posibilidad es dedicar un solo buffer circular grande por conexión. Este sistema también hace buen uso de la memoria mientras todas las conexiones estén muy cargadas, pero es deficiente si algunas de ellas no se encuentran en esta condición. Es claro entonces que mientras más simple sea la administración del buffer más simple se hace el problema. Por otra parte las técnicas de almacenamiento intermedio por flujo son auspiciosas con el fin de garantizar QoS [9], [10]. Luego, en base a todas las consideraciones anteriores se asume que la totalidad de las conexiones se encuentran muy cargadas y que se divide el ancho de banda promedio disponible.

De la discusión anterior se concluye que es posible analizar el problema como un caso de conexión como sigue. Considere que un flujo llega al nodo de destino desde cualquiera de los $n$ nodos fuente a lo largo de una secuencia compuesta por varios nodos intermedios. Los nodos en la ruta de la conexión están numerados por $1, \cdots, n$, y el nodo de origen está identificado por 0. El tiempo que se tarda en obtener servicio en cada nodo es finito y determinista. La tasa de envío desde la fuente se denomina $u_{0i}$ ($i = 1, \cdots, n$) y la tasa de servicio desde los nodos es $u_i$ ($i = 1, \cdots, n$). Junto con lo anterior se define

$$u_b = \min_i \{u_i \mid 0 \leq i \leq n\} \tag{1}$$

como la tasa de servicio del nodo cuello de botella.

Puesto que el modelo es discreto, el parámetro de tiempo $t$ que representa el tiempo continuo se reemplaza por el índice de paso $k$, el cual representa el tiempo de ida y vuelta RTT [11]. En el nodo cuello de botella la tasa de servicio se encuentra cerca de RTT [12], por lo tanto se considera que $u_b(k) = u_b(k+1)$ es una constante. En el momento marcado "NOW" de la Fig. 2 de [12]; al final de la $k$th época, todos los paquetes enviados en la época $k-1$ han sido reconocidos [11], [12]. Así es que los únicos paquetes no reconocidos son los enviados en la misma época y esto es lo mismo que hablar del número de paquetes independientes. Así este comportamiento puede ser aproximado por la tasa $u_0(k)$ de envío multiplicada por el intervalo de envío RTT($k$). Luego, se tiene que el modelo de este problema según [12] está dado por

$$q_b(k+1) = q_b(k) + u_0(k)\text{RTT}(k) - u_b(k)\text{RTT}(k), \tag{2}$$

donde $q_b(k)$ es la longitud de la cola del buffer en la época $k$th.

### III. Diseño del Controlador Discreto

Considere la entrada es la tasa de envío del nodo de origen y la salida es el nivel de la cola del cuello de botella. La Fig. 2 muestra el principio de control del sistema. En la Fig. 2 $S_i$ ($i = 0, \cdots, n$) denota nodo fuente; $D_i$ ($i = 0, \cdots, n$) denota el nodo destino; $B$ denota el nodo cuello de botella; $u_{0i}$ ($i = 0, \cdots, n$) denota la velocidad de envío, y $q_i$ ($i = 0, \cdots, n$) denota nivel de cola de buffer. El controlador $i$ ($i = 0, \cdots, n$) se diseña para controlar el sistema de red [13].

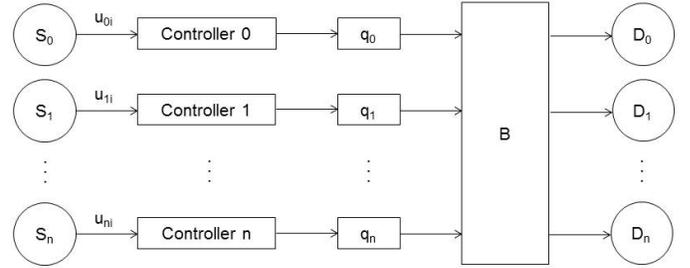

Fig. 2. Principio del sistema de control.

Tomando la transformada Z de (2), se obtiene

$$zq_b(z) = q_b(z) + u_0(z)\text{RTT}(z) - u_b(z)\text{RTT}(z). \tag{3}$$

Para cada conexión, TCP mantiene disponible el RTT, que es la mejor estimación actual del tiempo de ida y vuelta al destino en cuestión. Cuando se envía un segmento, se inicia un temporizador, tanto para ver cuánto tarda el acuse de recibo como para activar una retransmisión si tarda demasiado. Si el acuse de recibo regresa antes de que expire el temporizador, TCP mide cuánto tardó el acuse de recibo, $M$. Luego actualiza RTT de acuerdo con

$$\text{RTT}(k) = \alpha \text{RTT}(k-1) + (1-\alpha)M, \tag{4}$$

donde $\alpha$ es un factor de suavizado que determina cuánto peso se otorga al valor anterior. Típicamente $\alpha = 7/8$ [11].

Tomando la transformada Z de (4), se obtiene

$$\text{RTT}(z) = \alpha z^{-1}\text{RTT}(z) + (1-\alpha)M. \tag{5}$$

Luego

$$RTT(z) = \frac{z(1-\alpha)M}{z - \alpha}. \tag{6}$$

Sustituyendo (6) en (3), se obtiene

$$(z-1)q_b(z) = [u_0(z) - u_b(z)]\frac{M}{8 - 7z^{-1}}. \tag{7}$$

Sustituyendo $u_0(z) = u_b(z) + \lambda(z)$ de [12] en (6) se obtiene

$$(z-1)q_b(z) = \lambda(z)\frac{M}{8 - 7z^{-1}}. \tag{8}$$

Sea $\lambda(z) = f(z)S(z)$, donde $f(z)$ es una función racional de $\mathbb{Z}$ y $S(z)$ la función escalón; luego (7) se transforma en [12]

$$(z-1)q_b(z) = f(z)S(z)\frac{M}{8 - 7z^{-1}}, \tag{9}$$

que permite hallar la función de transferencia, $G(z)$, del sistema

$$G(z) = \frac{q_b(z)}{S(z)} = \frac{Mf(z)}{(z-1)(8 - 7z^{-1})}. \tag{10}$$

Sea $f(z) = \dfrac{c(z-1)(8z-7)}{(z+a)(z+b)}$, luego, de (9) se obtiene que

$$G(z) = \dfrac{Mcz}{(z+a)(z+b)}, \quad M, c \in \mathbb{R}. \tag{11}$$

En concordancia con la teoría de control lineal el sistema es estable si y solo si $|a|<1$, $|b|<1$.

De acuerdo con el teorema de la transformada Z, el error de paso en estado estacionario está dado por

$$q_b(\infty) = \lim_{z \to 1}(z-1)q_b(z) = \lim_{z \to 1}\dfrac{Mcz}{(z+a)(z+b)} = \dfrac{4}{5}Q, \tag{12}$$

donde Q es la capacidad máxima distribuida del buffer, el cual tiene un margen de representación 4/5 que no descarta paquetes cuando se produce el flujo de carga de entrada.

De (11) se tiene que

$$\lim_{z \to 1}\dfrac{Mcz}{(z+a)(z+b)} = \dfrac{4}{5}Q \Rightarrow c = \dfrac{4Q(1+a)(1+b)}{5M}. \tag{13}$$

Luego, dado que $\lambda(z) = f(z)S(z)$ [12], se tiene que

$$\lambda(z) = \dfrac{c(z-1)(8z-7)}{(1+a)(1+b)}\dfrac{z}{z-1}, \tag{14}$$

donde $z/(z-1)$ es la transformada Z de la función escalón.

Dado que $u_0(z) = u_b(z) + \lambda(z)$, de (14) se tiene que

$$u_0(z) = u_b(z) + \dfrac{c(z-1)(8z-7)}{(1+a)(1+b)}\dfrac{z}{z-1}. \tag{15}$$

Simplificado (15) se obtiene

$$u_0(z) = u_b(z) + \dfrac{cz(8z-7)}{(1+a)(1+b)}. \tag{16}$$

Expandiendo y factorizando (16) se obtiene

$$u_0(z) = u_b(z) + cz\left(\dfrac{a_1}{1+a} + \dfrac{a_2}{1+b}\right), \tag{17}$$

donde $a_1 = (-7-8a)/(b-a)$ y $a_2 = (-7-8a)/(a-b)$.

Sustituyendo (1) y (17) en (7) se obtiene

$$u_0(k) = u_b(k) + \dfrac{4Q(1+a)+(1+b)}{5M} + \left[\dfrac{-7-8a}{b-a}(-a)^k + \dfrac{-7-8b}{a-b}(-b)^k\right], \tag{18}$$

que es el algoritmo de control de la tasa de envío.

IV. Experimentos Numéricos

En la Tabla I se muestran los valores para los parámetros variables del sistema.

TABLA I
PARÁMETROS VARIABLES DEL SISTEMA

| Forma de Línea | $u_b(k)$ paquetes/ms | M ms | Q paquetes |
|---|---|---|---|
| Línea Sólida | 14,5 | 10 | 1000 |
| Línea punteada gruesa | 20 | 7,5 | 1000 |
| Línea punteada delgada | 27 | 5 | 1000 |

La Fig. 3 expone resultados para $a = -0,2$, $b = -0,1$ y $k = 0,8$ y en la Fig. 4 para $a = -0,5$, $b = -0,1$ y $k = 0,8$.

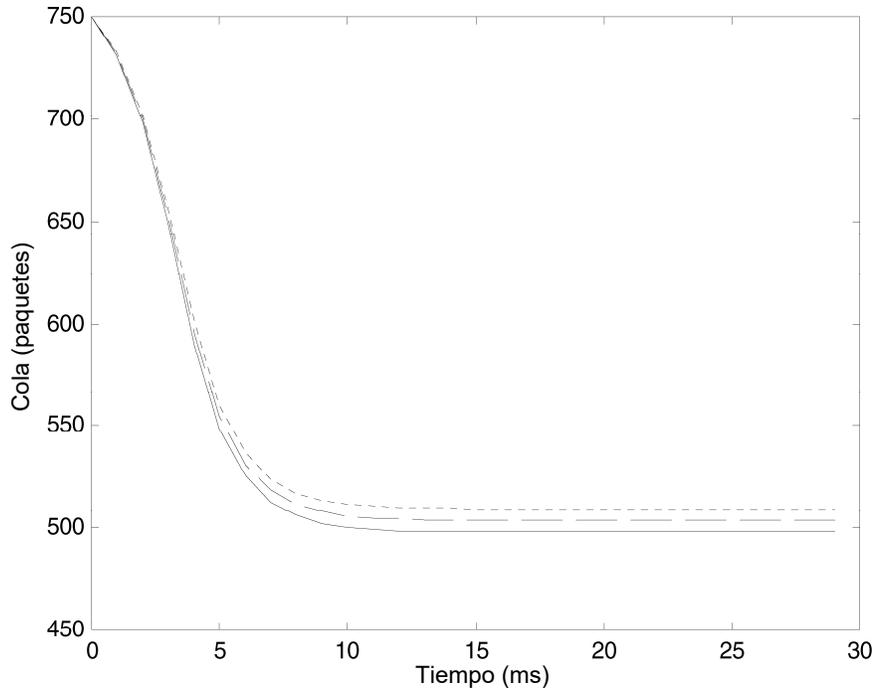

Fig. 3. Simulación considerando $a = -0,2$, $b = -0,1$ y $k = 0,8$.

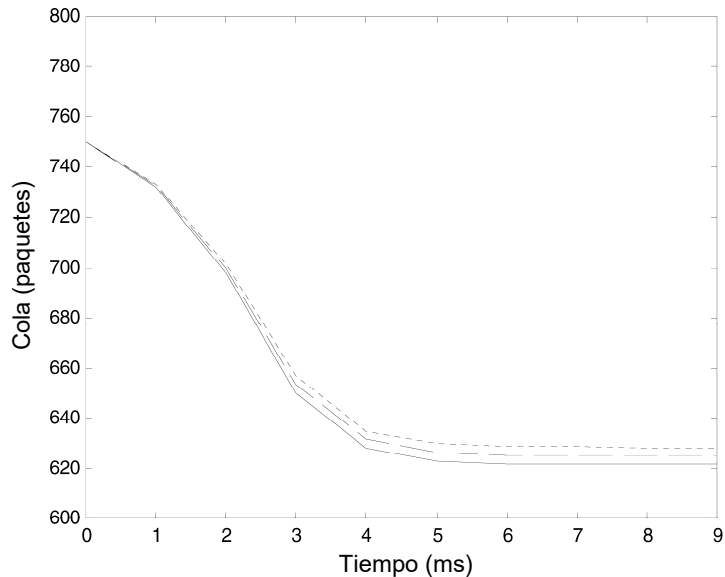

Fig. 4. Simulación considerando $a = -0,5$, $b = -0,1$ y $k = 0,8$.

A partir de los resultados de las simulaciones en la Fig. 3 y Fig. 4 se observa lo crítico que resulta una buena elección de $a$ y $b$, ello por la duración que puede llegar a tener el tiempo de ajuste frente el tiempo de estabilización del sistema.

Es claro a partir de los resultados de las simulaciones en la Fig. 3 y Fig.4 que se garantiza un rendimiento alto del sistema incluso si los parámetros varían, esto último debido al efecto de la ley de control aplicada.

La longitud de la cola del buffer se ve poco afectada por los parámetros variables y además se observa que se puede ajustar rápidamente a un valor determinado.

Finalmente se observa que el tiempo de estabilización del sistema depende de los parámetros $a$ y $b$.

## V. Conclusiones

Junto con el rápido desarrollo de las redes de computadoras sus requisitos de operación y funcionamiento se levantan cada vez más con exigencias de ingentes anchos de banda y tiempo muertos cada vez menores por no decir ininterrumpidos.

En todo lo anterior el control de la congestión desempeña un papel importante en el rendimiento que presente una red, y en este artículo se propone un nuevo y simple esquema de control de flujo basado en la teoría de control discreto y el análisis de la estabilidad y viabilidad del sistema.

Sobre la base de simulaciones se muestra que el algoritmo de control propuesto ajusta la tasa de envío y la cola del buffer rápida y eficazmente.

Además de lo anterior, la longitud final de la cola del búfer es poco afectada por parámetros variables y puede ser ajustada rápidamente a un valor dado.